\documentstyle[12pt,epsfig]{article} \input{psfig} 

\title{Possible evidence from laboratory measurements for a latitude and
longitude dependence of G}

\author{J.P.  Mbelek and M.  Lachi\`eze-Rey \\ Service d'Astrophysique, C.E.
Saclay \\ F-91191 Gif-sur-Yvette Cedex, France}

\begin{document} \maketitle \baselineskip=8mm

\begin{abstract} Stability arguments suggest that the Kaluza-Klein (KK) internal
scalar field, $\Phi$, should be coupled to some external fields.  An external
bulk real scalar field, $\psi$, minimally coupled to gravity is proved to be
satisfactory.  At low temperature, the coupling of $\psi$ to the electromagnetic
(EM) field allows $\Phi$ to be much stronger coupled to the EM field than in the
genuine five dimensional KK theory.  It is shown that the coupling of $\Phi$ to
the geomagnetic field may explain the observed dispersion in laboratory
measurements of the (effective) gravitational constant.  The analysis takes into
account the spatial variations of the geomagnetic field.  Except the high PTB
value, the predictions are found in good agreement with all of the experimental
data.  \end{abstract}

PACS numbers:  04.50.+h, 95.30.Sf, 06.20.Jr, 04.80.Cc

\section{Introduction} Although the methods and techniques have been greatly
improved since the late nineteenth century, the precision on the measurement of
the gravitational constant, G, is still the less accurate in comparison with the
other fundamental constants of nature \cite{Luo}.  Moreover, given the relative
uncertainties of most of the individual experiments (reaching about $10^{-4}$
for the most precise measurements), they show an incompatibility which leads to
an overall precision of only about 1 part in $10^{3}$ \cite{Chen}.  Thus the
current status of the $G$ terrestrial measurements (see \cite{Gillies}) implies
either an unknown source of errors (not taken into account in the published
uncertainties), or some new physics \cite{Melnikov}.  In the latter spirit, many
theories have been proposed as candidates for the unification of physics.  As
such, they involve a coupling between gravitation and electromagnetism
(hereafter GE coupling), as well as with other fields present.

The Kaluza-Klein theories \cite{Kaluza, Klein} have been among the first
attempts to unify electromagnetism and gravitation.  Although they are
disqualified in their original form, they constitute the prototype for many
present theories (in particular, those involving extra-dimensions).  In the
laboratory conditions for G-measurements, fundamental theories as well as KK
theories are well described as an \emph{effective} theory, under the form of the
Einstein and Maxwell equations conveniently modified, as described below.

This paper explores the possibility that the discrepancy between the results of
the G-measurements is the effect of the GE coupling, by comparing the available
data to the predictions of such an effective theory.  We do not pretend that the
effective theory adopted here (KK$\psi$, see below) is the only possibility.
Rather, we consider that it provides the simplest opportunity to confront the
idea of a GE coupling with real data.  The success of the fit suggests that this
may be the real explanation.

The simplest theories accounting for a GE coupling are those of Kaluza-Klein
\cite{Kaluza, Klein}, or slight modifications of them.  Their use corresponds to
the most economical way (the minimum of hypotheses) to test the hypothesis of GE
coupling.  As it is well known, such theories are effectively well described by
Einstein equations (with their correct Newtonian limit), where the newtonian
gravitational constant $G$ is replaced by $G_{eff}$, given below:  this
effective gravitational constant depends of the fifteenth degree of freedom,
${\hat{g}}_{44}$, of the (5-dimensional) bulk metric, which plays the role of a
(four dimensional) scalar field $\Phi$ (${\hat{g}}_{44} = - \,{\Phi}^{2}$ in the
Jordan-Fierz frame).

The genuine five dimensional Kaluza-Klein theories being subject to
instabilities (\cite{Witten, Salam, Sokolowski}), various authors
(\cite{Goldberger, Cline, Mbelek}) have suggested a more acceptable version
which includes an additional {\sl stabilizing} external bulk field:  here we
adopt the minimal hypothesis of a scalar field $\psi $ minimally coupled to
gravity.  In this theory (hereafter KK$\psi$), $G_{eff}$ varies with the
electromgnetic field, and thus in spacetime.  However, only variations with
respect to the cosmic distance or time have been investigated in the literature
hitherto.  But, since the geomagnetic field varies, with latitude and longitude,
and thus at the different sites of the $G$ measurements, The GE coupling implies
that the experiments in fact measure the distinct corresponding values of
$G_{eff}$, rather than an unique value of $G$.

Note also that this theory predicts a variation of the effective fine structure
constant $\alpha$ with the gravitational field, and thus with the cosmological
time.  In a companion paper (\cite{paper II}), we compare (with success) the
predicted evolution with astrophysical data concerning the distant quasars
(\cite{Webb, Murphy}).

In section 2 we recall the definition of the effective coupling constants, and
the effective Maxwell-Einstein equations, issued from the five dimensional
compactified KK theory stabilized by a minimally coupled bulk scalar field.  In
section 3, we calculate the vacuum solutions on Earth, in the weak field limit,
taking into account the geomagnetic field.  In section 4, we confront the
predicted values of $G_{eff}$ with the laboratory measurements.  In section 5,
we discuss the consistency of our results with respect to the orbital motion of
the LAGEOS satellite, the Moon and planets of the solar system, as well as the
binary pulsar PSR$1913+16$.

\section{Theoretical background}

An argument initially from Landau and Lifshitz \cite{Landau} may be applied to
the pure Kaluza-Klein (KK) action (\cite{Mbelek}):  the negative sign of the
kinetic term of the KK internal scalar field leads to inescapable instability.
Stabilization may however be obtained if an external field is present
(\cite{Witten}, \cite{Salam}, \cite{Mbelek}), and we assume here a version
KK$\psi$ of the KK theory which includes an external bulk scalar field minimally
coupled to gravity.  After dimensional reduction ($\alpha = 0, 1, 2, 3$), this
bulk field reduces to a four dimensional scalar field $\psi = \psi (x^{\alpha})$
and, in the Jordan-Fierz frame, the low energy effective action takes the form
(up to a total divergence) $$ S = - \,\int \sqrt{-g} \,\, [
\,\frac{c^{4}}{16\pi} \,\frac{\Phi}{G} \,R \,\,+ \,\,\frac{1}{4}
\,\,{\varepsilon}_{0} \,{\Phi}^{3} \,F_{\alpha\beta} \,F^{\alpha\beta} \,\,+
\,\,\frac{c^{4}}{4\pi G} \,\frac{{\partial}_{\alpha} \Phi \,{\partial}^{\alpha}
\Phi}{\Phi} \,] \,d^{4}x $$ \begin{equation} \label{KK action and Psi J-F frame}
\,+ \,\int \sqrt{-g} \,\,\Phi \,[ \,\frac{1}{2} \,{\partial}_{\alpha} \psi
\,\,{\partial}^{\alpha} \psi \,\,- \,\,U \,\,- \,\,J \psi \,] \,d^{4}x,
\end{equation} where $A^{\alpha}$ is the potential 4-vector of the
electromagnetic field, $F_{\alpha\beta} = {\partial}_{\alpha} \,A_{\beta} \,-
\,{\partial}_{\beta} \,A_{\alpha}$ the electromagnetic field strength tensor,
$U$ the self-interaction potential of $\psi$ and $J$ its source term.  Following
Lichnerowicz \cite{Lichnerowicz}, we interpret the quantity \begin{equation}
\label{G eff} G_{eff} = \frac{G}{\Phi} \end{equation} of the Einstein-Hilbert
term, and the factor \begin{equation} \label{varepsilon eff}
{\varepsilon}_{0eff} = {\varepsilon}_{0} \,{\Phi}^{3} \end{equation} of the
Maxwell term respectively as the effective gravitational "constant" and the
effective vacuum dielectric permittivity.  The effective vacuum magnetic
permeability reads ${\mu}_{0eff} = {\mu}_{0}/{\Phi}^{3}$, so that the velocity
of light in vacuum remains a true universal constant.  Both terms depend solely
on the local (for local physics) or global (at cosmological scale) value of the
KK scalar field $\Phi$, assumed to be positive defined.

The source term of the $\psi$-field, $J$, includes the contributions of the
ordinary matter (other than the scalar fields $\psi$ and $\Phi$), of the
electromagnetic field and of the internal scalar field $\Phi$.  For each, the
coupling is defined by a function (temperature dependent, as for the potential
$U$) $f_{X} = f_{X}(\psi\,,\,\Phi)$, where the subscript $X$ stands for
"matter", "EM" and "$\Phi$".  In order to recover the Einstein-Maxwell equations
in the weak fields limit, these three functions are subject to the conditions:
$f_{EM}(v\,,\,1) =f_{matter}(v\,,\,1) = f_{\Phi}(v\,,\,1) = 0$, where $v$
denotes the vacuum expectation value (VEV) of the $\psi$-field.

The contributions of matter and $\Phi$ are proportional to the traces of their
respective energy-momentum tensors.  Since the energy-momentum tensor of the
electromagnetic field is traceless, a contribution of the form
${\varepsilon}_{0}\,f_{EM} \,F_{\alpha\beta} \,F^{\alpha\beta}$ accounts for the
coupling with it.  The fit of our model to the data (see below) shows that
$\frac{\partial f_{EM}}{\partial\Phi}\,(v\,,\,1)\,v \gg 4\pi G/c^{4}$, as it can
be expected near the vacuum at low temperature.  Thus, we will not take the
latter term into account.  However, we may suspect that $\frac{\partial
f_{EM}}{\partial\Phi}\,(v\,,\,1)\,v \leq 4\pi G/c^{4}$ at high temperature,
which may have consequences in some astrophysical conditions (see below).

Applying the least action principle to the action (\ref{KK action and Psi J-F
frame}) yields :

\begin{itemize}

\item the generalized Einstein equations \begin{equation} \label{Einstein eq}
R_{\alpha\beta} \,- \,\frac{1}{2} \,R\,g_{\alpha\beta} = \frac{8\pi
G_{eff}}{c^{4}} \,[ \,T^{(EM)}_{\alpha\beta} \,+ \,T^{(\Phi)}_{\alpha\beta} \,+
\,T^{(\psi)}_{\alpha\beta} \,], \end{equation} where \begin{equation}
\label{effective EM energy-momentum tensor} T^{(EM)}_{\alpha\beta} =
{\varepsilon}_{0eff} \,( \,- \,F^{\gamma}_{\alpha}\,F_{\beta\gamma} \,+
\,\frac{1}{4} \,F_{\gamma\delta}\,F^{\gamma\delta} \,g_{\alpha\beta} \,),
\end{equation} \begin{equation} \label{effective Phi energy-momentum tensor}
T^{(\Phi)}_{\alpha\beta} = \frac{c^{4}}{8\pi G} \,(
\,{\nabla}_{\alpha}\,{\nabla}_{\beta}\,\Phi \,- \,g_{\alpha\beta}
\,{\nabla}_{\nu}\,{\nabla}^{\nu}\,\Phi \,) \end{equation} and \begin{equation}
\label{effective psi energy-momentum tensor} T^{(\psi)}_{\alpha\beta} := \Phi
\,[ \,{\partial}_{\alpha}\,\psi \,{\partial}_{\beta}\,\psi \,\,- \,\,(
\,\frac{1}{2} \,{\partial}_{\gamma}\,\psi \,{\partial}^{\gamma}\,\psi \,\,-
\,\,U \,\,- \,\,J \psi \,) \,g_{\alpha\beta} \,] \end{equation}

\item the generalized Maxwell equations \begin{equation} \label{1st group
Maxwell eq} {\nabla}_{\gamma}\,F_{\alpha\beta} \,+
\,{\nabla}_{\beta}\,F_{\gamma\alpha} \,+ \,{\nabla}_{\alpha}\,F_{\beta\gamma} =
0 \end{equation} and \begin{equation} \label{2nd group Maxwell eq}
{\nabla}^{\alpha} \,( \,{\varepsilon}_{0eff}\,F_{\alpha\beta} \,) = 0,
\end{equation}

\item and the scalar fields equations \begin{equation} \label{ext.  scalar field
eq} {\nabla}_{\nu} {\nabla}^{\nu} {\psi} = - \,\,J \,\,- \,\,\frac{\partial J}
{\partial {\psi}} \,\psi \,\,- \,\frac{\partial U} {\partial {\psi}}
\end{equation} and \begin{equation} \label{KK scalar eq}
{\nabla}_{\nu}\,{\nabla}^{\nu}\,\Phi = - \,\frac{4\pi G}{c^{4}}
\,{\varepsilon}_{0} \,F_{\alpha\beta}\,F^{\alpha\beta}\,{\Phi}^{3} \,\,+
\,\,U\,\Phi \,\,+ \,\,J \psi\,\Phi \,\,+ \,\,\frac{\partial J} {\partial {\Phi}}
\,{\Phi}^{2} \,\psi \,\,- \,\,\frac{1}{2} \,( \,{\partial}_{\alpha} \psi
\,\,{\partial}^{\alpha} \psi \,) \,\Phi, \end{equation}

where the symbol ${\nabla}_{\nu}$ stands for the Riemannian covariant
derivative.  Clearly, $T^{(EM)}_{\alpha\beta}$ and $T^{(\Phi)}_{\alpha\beta}$
define respectively an effective energy-momentum tensor for the electromagnetic
field in the presence of the KK scalar field and an effective energy-momentum
tensor for the latter itself.  Relations (\ref{Einstein eq}-\ref{2nd group
Maxwell eq}) are formally the same as the Einstein-Maxwell ones, but with the
additional contribution of the KK scalar as a matter source and the replacement
of $G$ and ${\varepsilon}_{0}$ by their respective effective values.

\end{itemize}

\section{Vacuum solutions in the presence of a dipolar magnetic field}

Since we are in weak field conditions (we look for small deviations from
Newtonian physics), we only keep first order terms.  Thus, we neglect the
excitations of $\Phi$ and $\psi$ with respect to their respective VEV's 1 and
$v$.  Also, the energy density of the $\psi$-field must be lower than that of
the magnetic field.\\ Let us study the {\sl spatial} variation of $\psi$ out of
the fields' source, but in presence of a static dipolar magnetic field, $\vec{B}
= \vec{B}(r, \varphi, \theta)$.  We denote $r$, $\varphi$ and $\theta$
respectively the radius from the centre, the azimuth angle and the colatitude.
Thus, writing $\psi = \psi (r, \varphi, \theta)$ and $\Phi = \Phi (r, \varphi,
\theta)$, and taking into account that $\frac{\partial U}{\partial \psi} (v) =
0$ (definition of the VEV), equations (\ref{ext.  scalar field eq}) and (\ref{KK
scalar eq}) simplify respectively as \begin{equation} \label{ext.  scalar field
eq stat.  1st approx} \frac{1}{r^{2}} \,[ \,\frac{\partial}{\partial r} \,(
\,r^{2} \,\frac{\partial \psi}{\partial r} \,) \,+ \,\frac{1}{\sin{\theta}}
\,\frac{\partial}{\partial \theta} \,(\, \sin{\theta} \,\frac{\partial
\psi}{\partial \theta} \,) \,+ \, \frac{1}{\sin^{2}{\theta}}
\,\frac{{\partial}^{2} \psi}{\partial {\varphi}^{2}} \,] = 2 \,\frac{\partial
f_{EM}}{\partial \psi}\,(v\,,\,1)\,v \,\frac{B^{2}}{{\mu}_{0}} \end{equation}
and \begin{equation} \label{stabilized KK scalar eq stat.  1st approx}
\frac{1}{r^{2}} \,[ \,\frac{\partial}{\partial r} \,( \,r^{2} \,\frac{\partial
\Phi}{\partial r} \,) \,+ \,\frac{1}{\sin{\theta}} \,\frac{\partial}{\partial
\theta} \,(\, \sin{\theta} \,\frac{\partial \Phi}{\partial \theta} \,) \,+ \,
\frac{1}{\sin^{2}{\theta}} \,\frac{{\partial}^{2} \Phi}{\partial {\varphi}^{2}}
\,] = - \,2 \,\frac{\partial f_{EM}}{\partial \Phi}\,(v\,,\,1)\,v
\,\frac{B^{2}}{{\mu}_{0}}, \end{equation} where we have dropped the pure
gravitational constant $4\pi G/c^{4}$ with respect to $v\,\partial
f_{EM}/\partial \Phi$, as indicated above.

We consider a dipolar magnetic field:  $\vec{B} = \vec{\nabla}\,V$.  For our
purpose, it is sufficient to limit the expansion of the scalar potential, $V$,
to the terms of the Legendre function of degree one ($n = 1$) and order one ($m
= 1$).  Hence $V = (a^{3}/r^{2}) \,[ \,g^{0}_{1}\,\cos{\theta} \,+
\,g^{1}_{1}\,\sin{\theta}\,\cos{\varphi} \,+
\,h^{1}_{1}\,\sin{\theta}\,\sin{\varphi} \,]$, where $g^{0}_{1}$, $g^{1}_{1}$
and $h^{1}_{1}$ are the relevant Gauss coefficients, $a$ is the Earth's radius
and $M = \frac{4\pi}{{\mu}_{0}} \,a^{3}\,\sqrt{(g^{0}_{1})^{2} \,+
\,(g^{1}_{1})^{2} \,+ \,(h^{1}_{1})^{2}}$ denotes its magnetic moment (see
\cite{IGRF coefficients, Rishbeth}).  Setting $\cos{{\varphi}_{1}} =
-\,g^{1}_{1}/\sqrt{(g^{1}_{1})^{2} \,+ \,(h^{1}_{1})^{2}}$, $\sin{{\varphi}_{1}}
= h^{1}_{1}/\sqrt{(g^{1}_{1})^{2} \,+ \,(h^{1}_{1})^{2}}$ and $\tan{\lambda} =
g^{0}_{1}/\sqrt{(g^{1}_{1})^{2} + (h^{1}_{1})^{2}}$, the solutions of equation
(\ref{stabilized KK scalar eq stat.  1st approx}) then reads \begin{equation}
\label{stabilized KK scalar sol.  stat.  1st approx} \Phi = 1 \,\,-
\,\,k(r)\,x(\theta, \varphi), \end{equation} where we have set \begin{equation}
\label{def.  radial function k} k(r) = \frac{1}{{\mu}_{0}} \,\frac{\partial
f_{EM}}{\partial \Phi}\,(v\,,\,1)\,v \,( \,\frac{{\mu}_{0}\,M}{4\pi\,r^{2}}
\,)^{2} \end{equation} and $$ x(\theta, \varphi) = \cos^{2}{\theta} \,\,+
\,\,\epsilon(\theta, \varphi) $$ \begin{equation} \label{def.  mixed variable x}
= \cos^{2}{\theta} \,\,+
\,\,\cot^{2}{\lambda}\,\sin^{2}{\theta}\,\cos^{2}{(\,\varphi \,\,+
\,\,{\varphi}_{1}\,)} \,\,- \,\,\cot{\lambda}\,\sin{2\theta}\,\cos{(\,\varphi
\,+ \,{\varphi}_{1}\,)} \end{equation} and similarly for $\psi$ by making the
substitution $\frac{\partial f_{EM}}{\partial \Phi} \rightarrow -
\,\frac{\partial f_{EM}}{\partial \psi}$ in relation (\ref{def.  radial function
k}).  Thence, one derives the expression of $G_{eff}(r, \theta, \varphi)$ by
inserting the solution (\ref{stabilized KK scalar sol.  stat.  1st approx})
above in relation (\ref{G eff}).  Thus, since $\frac{\partial f_{EM}}{\partial
\Phi}\,(v\,,\,1)\,v > 0$ and the variable $x$ turns out to be positive at any
position in space, it follows that the effective gravitational constant
$G_{eff}$ will always be greater than the true gravitational constant, $G$.
Whence the prediction of an upward bias in the laboratory measurements of G.

\section{Comparison with laboratory measurements}

Because of various uncontrolled systematic errors, the data published by the
different laboratories have different precisions.  In the following, we test two
hypotheses with respect to these results:  H0 = Hypothesis of a constant G ($\nu
= n \,- \,1$) and H1 = Hypothesis of an effective G ($\nu = n \,- \,2$).  Here
$\nu$ denotes the number of degrees of freedom, $n$ is the number of data
points, and we have 1 or 2 parameters in the fit.

There are presently almost 45 results of measurements G published since 1942
(see {\sl e.g.}, \cite{Gillies}, Table 2, pp.  168 and 169).  We exclude from
the present study the mine measurements because of the too numerous uncontrolled
systematic biases involved.  The " accepted " values are presently $G = 6.67259
\pm 0.00085~10^{-11}$ (CODATA 86, \cite{CODATA 86}) and $G = 6.670 \pm
0.010~10^{-11}$ (CODATA 2000, \cite{CODATA 00}) in MKS unit.  A fitting of the
45 data with these values give respectively $\chi^{2}_{\nu} = 145.17$ and
$\chi^{2}_{\nu} = 213.25$ ($\chi^{2}_{\nu} = \chi^{2}/\nu$, where $\nu$ denotes
the degrees of freedom).  If we forget the accepted value and try a best fit,
assuming an arbitrary constant value of $G$, we obtain $G = 6.6741~10^{-11}$ SI
with $\chi^{2}_{\nu} = 127.04$.

The more discordant laboratory measurement (high PTB value \cite{Braunschweig
95}) is controversial.  If we discard it, the previous fits lead to
$\chi^{2}_{\nu} = 11.128$ (CODATA 86), $\chi^{2}_{\nu} = 62.498$ (CODATA 2000)
and $\chi^{2}_{\nu} = 2.341$ (free value).  Since it seems now certain that this
high PTB value \cite{Braunschweig 95} suffers from some systematic error (see
\cite{Sevres 01}, for more details), we remove it for our analysis (if we keep
it, our model is still more favored).  Note that the strongest contributions to
the $\chi^{2}_{\nu}$ then come from the BIPM 2001 \cite{Sevres 01} and the HUST
\cite{Wuhan} measurements.  Thus, unless there are some experimental systematic
errors presently not understood, these experiments do not measure the same
quantity.  In the framework of the theory proposed here, they measure $G_{eff}$.
Because of the GE coupling, $G_{eff}$ should depend on the geomagnetic field at
the laboratory position.

First, we select a subset of results with a good precision and which does not
suffer any significant systematic error:  we include only those points with
relative uncertainty $\delta G_{lab}/G_{lab} \leq 10^{-3}$.  Also, we demand a
short measuring time ($\Delta t < 200$~s), to minimize the possible biases due
to the time variations of the geomagnetic field.  This gives the sample S1, with
17 points, shown in figure 1.  We fit these data using the IGRF 2000 Gauss
coefficients\footnote{The IGRF coefficients are given for time intervals of five
years.  Hence, for a more precise fitting, one should use the most suitable IGRF
coefficients for a given laboratory value according to the years the
measurements were carried out.  Of course, the necessity of doing so depends on
the precision reached in the laboratory G measurements.  Nevertheless, we have
also computed the variable $x(L, l)$ using the IGRF 1965 Gauss coefficients
($g^{0}_{1} = - \,0.30339$, $g^{1}_{1} = - \,0.02123$ and $h^{1}_{1} = 0.05758$
in Gauss).  However, no significant changes to our conclusions were found.}
($g^{0}_{1} = - \,0.31543$, $g^{1}_{1} = - \,0.02298$ and $h^{1}_{1} = 0.05922$
in Gauss).  We obtain (Figure 1, Table 2) \begin{equation} \label{1/Glab vs x}
\frac{1}{10^{11}~G_{eff}} = ( \,0.149933 \pm 0.000017 \,) \,\,- \,\,(
\,0.0001514 \pm 0.0000262 \,) \,x(L, l), \end{equation} in MKS units, with
${\chi}^{2}_{\nu} = 1.327$.  This gives \begin{equation} \label{G laboratory
data} G = ( \,6.6696 \pm 0.0008 \,)~10^{-11}~m^{3}~kg^{-1}~s^{-2},
\end{equation} that we retain as a {\sl true} gravitational constant.  The
relative uncertainty is only 1 part in $10^{4}$:  the major part of the
differences between the laboratory measurements was generated by the predicted
variation of $ G_{eff}$ with the magnetic field.  Further, adjusting to the same
set the mean value $M = 7.87~10^{22}$ A m$^{2}$ (in the time interval spanning
from $1942$ to $2001$), it follows \begin{equation} \label{coupling constant}
\frac{\partial f_{EM}}{\partial \Phi}\,(v\,,\,1)\,v = ( \,5.44 \pm 0.66
\,)~10^{-6} \,fm\,\,Tev^{-1}, \end{equation} that we retain too.  We observe
also that the HUST value (the lowest most precise measured value of $G$) is
perfectly fitted:  it differs from other values because of the proximity of this
laboratory to the equator.

Then, we fit the whole sample (excluding the PTB value, as stated above).  We
obtain the set of the 44 measurements given in figure 2:  the best fit leads to
\begin{equation} \label{1/Glab vs x whole} \frac{1}{10^{11}~G_{eff}} = (
\,0.149929 \pm 0.000017 \,) \,\,- \,\,( \,0.0001509 \pm 0.0000252 \,) \,x(L, l)
\end{equation} in MKS units.  It gives $\chi^2_{\nu} = 1.669$, to be compared to
$\chi^2_{\nu} = 2.255$ for the best fit assuming a constant $G$.  These results
are summarized in Table 2.  In order to check the relevance of our result (which
involves two free parameters rather than one), we apply the F test (Fisher law).
This yields $F_{\chi} = \frac{\Delta \chi^2}{\chi^2_{\nu}} = 16.09$, which
indicates that, independently of the number of parameters, our fit is better
with a significance level greater than $99.9\%$ \cite{Bevington}.

\begin{tabular}{|c|c|c|c|} \hline \emph{Location [reference]} & \emph{Latitude
($^{\circ}$)} & \emph{Longitude ($^{\circ}$)} & \emph{G$_{lab}$ \,($10^{-11}
\,m^{3} \,kg^{-1} \,s^{-2}$)} \\ \hline Lower Hutt (MSL) \cite{Lower Hutt 99,
Lower Hutt 95} & -41.2 & 174.9 & 6.6742 $\pm$ 0.0007 \\ & & & 6.6746 $\pm$
0.0010 \\ \hline Wuhan (HUST) \cite{Wuhan} & 30.6 & 106.88 & 6.6699 $\pm$ 0.0007
\\ \hline Los Alamos \cite{Los Alamos} & 35.88 & -106.38 & 6.6740 $\pm$ 0.0007
\\ \hline Gaithersburg (NBS) \cite{Gaithersburg 82, Gaithersburg 42} & 38.9 &
-77.02 & 6.6726 $\pm$ 0.0005 \\ & & & 6.6720 $\pm$ 0.0041 \\ \hline Boulder
(JILA) \cite{Boulder} & 40 & -105.27 & 6.6873 $\pm$ 0.0094 \\ \hline Gigerwald
lake \cite{Gigerwald lake 95, Gigerwald lake 94} & 46.917 & 9.4 & 6.669 $\pm$
0.005 (at 112 m) \\ & & & 6.678 $\pm$ 0.007 (at 88 m) \\ & & & 6.6700 $\pm$
0.0054 \\ \hline Zurich \cite{Zurich 98, Zurich 99} & 47.4 & 8.53 & 6.6754 $\pm$
0.0005 $\pm$ 0.0015 \\ & & & 6.6749 $\pm$ 0.0014 \\ \hline Budapest
\cite{Budapest} & 47.5 & 19.07 & 6.670 $\pm$ 0.008 \\ \hline Seattle
\cite{Seattle} & 47.63 & - 122.33 & 6.674215 $\pm$ 0.000092 \\ \hline Sevres
(BIPM) \cite{Sevres 01, Sevres 99} & 48.8 & 2.13 & 6.67559 $\pm$ 0.00027 \\ & &
& 6.683 $\pm$ 0.011 \\ \hline Fribourg \cite{Fribourg} & 46.8 & 7.15 & 6.6704
$\pm$ 0.0048 (Oct.  84) \\ & & & 6.6735 $\pm$ 0.0068 (Nov.  84) \\ & & & 6.6740
$\pm$ 0.0053 (Dec.  84) \\ & & & 6.6722 $\pm$ 0.0051 (Feb.  85) \\ \hline
Magny-les-Hameaux \cite{Magny-les-Hameaux} & 49 & 2 & 6.673 $\pm$ 0.003 \\
\hline Wuppertal \cite{Wuppertal} & 51.27 & 7.15 & 6.6735 $\pm$ 0.0011 $\pm$
0.0026 \\ \hline Braunschweig (PTB) \cite{Braunschweig 95, Braunschweig 87} &
52.28 & 10.53 & 6.71540 $\pm$ 0.00056 \\ & & & 6.667 $\pm$ 0.005 \\ \hline
Moscow \cite{Moscow 98, Moscow 79} & 55.1 & 38.85 & 6.6729 $\pm$ 0.0005 \\ & & &
6.6745 $\pm$ 0.0008 \\ \hline Dye 3, Greenland \cite{Dye 3} & 65.19 & -43.82 &
6.6726 $\pm$ 0.0027 \\ \hline Lake Brasimone \cite{lake Brasimone} & 43.75 &
11.58 & 6.688 $\pm$ 0.011 \\ \hline \end{tabular}\\\\Table 1 :  Results of the
most precise laboratory measurements of G published during the last sixty years
and location of the laboratories.\\

\begin{tabular}{|c|c|c|} \hline Sample & H0 & H1 \\ \hline S1 & ~ &~ \\ 17
     points & $\chi^{2}_{\nu}$ = 3.607 (best fit) & \\ & $\chi^{2}_{\nu}$ =
     21.523 (mean of CODATA 86) & $\chi^{2}_{\nu}$ = 1.327\\ (Fig.1)&
     $\chi^{2}_{\nu}$ = 141.46 (mean of CODATA 2000) & \\ \hline Whole &
     $\chi^{2}_{\nu}$ = 2.255 (best fit) & \\ \cite{Lower Hutt 99} -
     \cite{others}& $\chi^{2}_{\nu}$ = 11.128 (mean of CODATA 86) & \\ 44 points
     & $\chi^{2}_{\nu}$ = 62.498 (mean of CODATA 2000) & $\chi^{2}_{\nu}$ =
     1.669 \\(Fig.2) &~ &~ \\ \hline \end{tabular}\\\\\\Table 2 :  Reduced
     $\chi^{2}$ for the two different hypothesis H0 (Hypothesis of a constant
     $G$) and H1 ((Hypothesis of an effective G), and different samples S1 and
     whole (except the high PTB value \cite{Braunschweig 95}, see text).\\

\begin{figure} \centerline{\epsfxsize=12cm \epsfbox{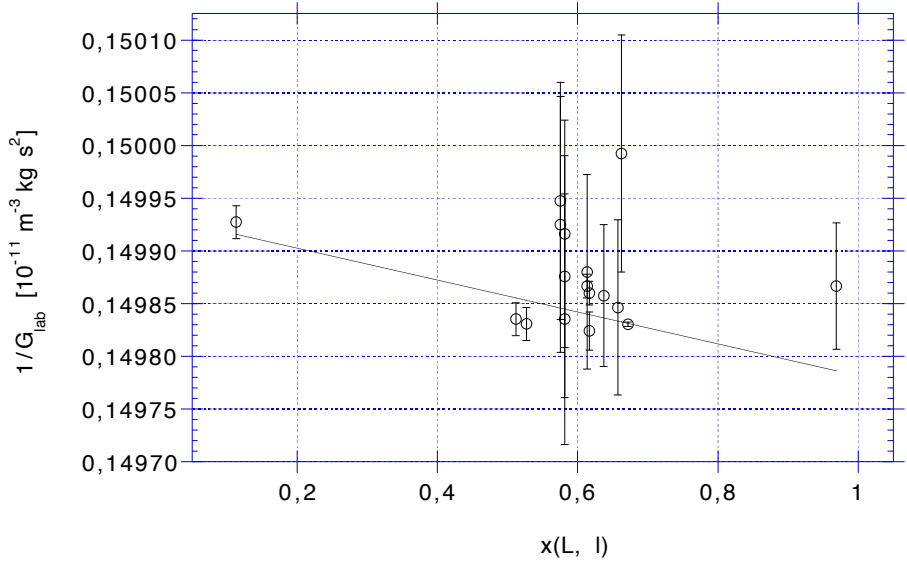}}
\caption{Laboratory measurements with relative uncertainty $\frac{\delta
G_{lab}}{G_{lab}} < 10^{-3}$ and measuring time $\Delta t < 200~s$ (sample S1,
17 points \cite{Lower Hutt 99}, \cite{Wuhan} - \cite{Gaithersburg 42},
\cite{Gigerwald lake 94}, \cite{Seattle}, \cite{Fribourg} - \cite{Wuppertal},
\cite{Braunschweig 87} - \cite{Dye 3}).  The line indicates the best fit
$G_{lab}$ versus the mixed variable $x$ ($\chi^{2}_{\nu} = 1.327$).  Assuming a
constant $G$ would yield a bad fit to the data ($\chi^{2}_{\nu} = 3.607$),
mostly because of the HUST value.}  \end{figure}

\begin{figure} \centerline{\epsfxsize=12cm \epsfbox{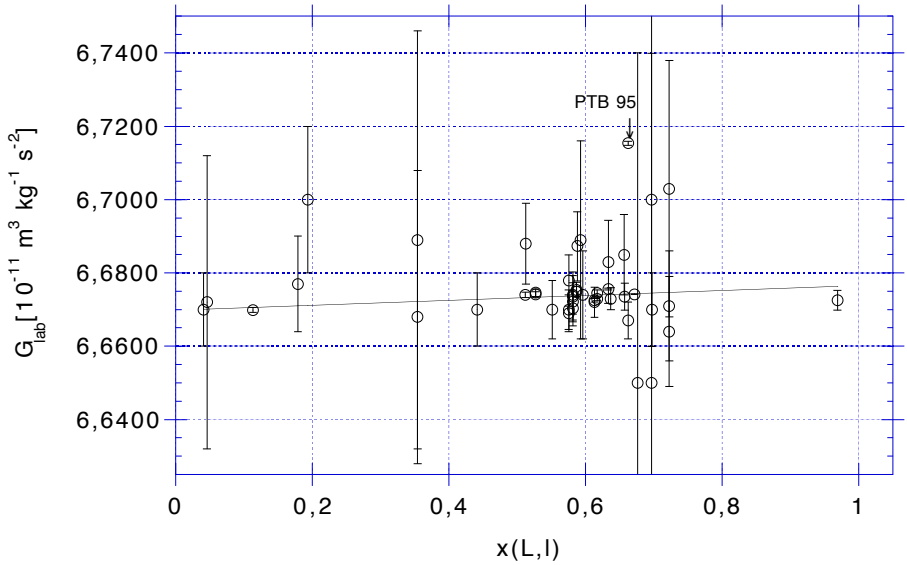}}
\caption{$G_{lab}$ versus $x$ (whole sample plus the PTB 95 value, 45 points
\cite{Lower Hutt 95} - \cite{others}).}  \end{figure}

\section{Discussion and conclusion}

We have not taken into account the temporal variation of the geomagnetic field.
All the data considered in this paper are averaged on time.  Nevertheless, for
those which would not be averaged on time, one may expect small time variations
of $G_{eff}$ both with the Sq and with the L field disturbances of the
geomagnetic field.  Now, periodic variations of the gravitational "constant"
with the lunar or diurnal period have yet been pointed out in the literature
(see \cite{Izmailov, Los Alamos, Boulder, Moscow 98}).  Although it is presently
believed that they are related to tides, the explanation could be this temporal
variation.  We notice that the G measurements of \cite{Fribourg} are consistent
with an annual variation.

Besides, a recent study \cite{Lopes} shows that \emph{ helioseismology} seems to
favor a low value of $G$, close to the HUST value.  This is in accordance with
our predictions (see section 3 and relations (\ref{1/Glab vs x}) and (\ref{G
laboratory data})), since one expects the effective coupling of the $\Phi$-field
to the EM field to decrease towards its lowest value as the temperature
increases.  Hence, one gets a good agreement between our prediction (\ref{G
laboratory data}) and the helioseismic data, on account of the high temperature
met in the core of the Sun.

To test our proposal of a spatial dependence of $G_{eff}$, we call for a
complete covering of the Earth (in particular in the south hemisphere) and at
different latitudes.  On the basis the current available data, the present study
allows to make some predictions.  First, according to the authors \cite{Lopes},
the Sudbury Neutrino Observatory (SNO:  $L = 46^{\circ}$ 29' N and $l =
80^{\circ}$ 59' O) \cite{Sudbury} should constitute a promising means of
determining G with a quite good accuracy.  Hence, since the measured quantity is
actually the effective gravitational constant, the value predicted at SNO by the
relation (\ref{1/Glab vs x}) is $G_{SNO} = ( \,6.6742 \pm 0.0009 \,)~10^{-11}$
m$^{3}$ kg$^{-1}$ s$^{-2}$, that is practically the same as at the MSL
\cite{Lower Hutt 99}.  In addition, it would be of interest to test also our
predicted value at the surface of the poles, that is:  $G_{poles} = ( \,6.6763
\pm 0.0010 \,)~10^{-11}$ in MKS units.  At the equator, one should get a
dependence with respect to the longitude (since the magnetic and geographic
poles do not coincide).

At the orbit of the Moon, the relative deviation, $(G_{eff} - G)/G$, to the true
gravitational constant is predicted to be as small as $1.7~10^{-13}$ which is
consistent with lunar laser ranging (see, \cite{LLR}).  In the Solar System, the
relevant magnetic field is the dipolar field of the quiet Sun.  Since the
coupling constant of the $\Phi$-field to the EM field is much weaker within the
Sun than on Earth, one finds at the orbital radius of Mercury $(G_{eff} - G)/G <
10^{-6}$ and much smaller beyond, decreasing as $1/r^{4}$.  Thus, at the orbital
radius of Neptune $(G_{eff} - G)/G$ drops to less than $10^{-12}$.  Taking into
account the overall planetary constraints on $GM_{\odot}$ \cite{Planetary G},
the accord between the proposed model and observation is still acceptable.

For artificial satellites in quasi-circular orbits, the appropriate quantity for
comparison with observational data is $(G_{eff} - G')/G'$ when the analysis is
based only on the satellite data (see appendices A and B), where we have set $G'
= G \,[ \,1 \,+ \,\frac{k(r)}{3} \,( \,13 \,+ \,\frac{3}{2} \,( \,\frac{a}{r}
\,)^{2} \,J_{2} \,) \,]$ and $J_{2}$ is the true Earth quadrupole moment
coefficient.  Referring to the orbital motion of the LAGEOS satellite, with
semimajor axis equal to $12~270~$km, inclination $i = 109.94^{\circ}$ and
eccentricity $e = 0.004$, one finds the maximum deviations $(G_{eff} - G')/G'
\simeq 1.9~10^{-8}$ (whereas $(G_{eff} - G)/G \simeq (G' - G)/G \simeq
2~10^{-5}$), consistent with the constraint $\mid \alpha \mid \,< 10^{-5} -
5~10^{-8}$ on the Yukawa coupling constant $\alpha$ (see \cite{LAGEOS}, figure
3.2a, p.  99 and section 6.7).\\ We also looked for a possible relative
deviation to the true gravitational constant induced by the strong magnetic
fields of \emph{pulsars}.  We found this effect quite negligible, of the order
$10^{-7}$ in the case of the binary pulsar PSR$1913+16$, for which observation
yields $G_{eff} = G_{N}\,(1.00^{+0.14}_{-0.11})$ \cite{PSR1913+16}.  The
deviation is tiny because of the small radius of the pulsar and of its companion
as compared to their respective orbital radius around the center of mass.

Thus we conclude that, apart from systematic errors that need to be corrected
({\it e.g.}, by applying some prescriptions like that pointed out by Kuroda
\cite{Kuroda} for the swinging pendulum method), the only possibility to
reconcile the published values of $G$ is to consider a dependence on the
latitude and longitude, of the type proposed here.  In particular, if all
present systematic errors could be removed in the future, we predict $G_{lab} =
( \,6.6742 \pm 0.0009 \,)~10^{-11}$ m$^{3}$ kg$^{-1}$ at the PTB laboratory,
that is the same value as that predicted at SNO and the current MSL.

Up to now, a lot of attention had been paid to the dependence of $G$ on cosmic
time or radial distance only.  But the dependence on latitude and longitude,
that we examine here, has not been taken into account.  More precise
measurements ({\sl e.g.}, the SEE project \cite{SEE}) and further analyses
taking into account higher harmonics and the various kind of changes of the
geomagnetic field should bring more support to our claim.

\section{Appendix A} The generalized Einstein equations (\ref{Einstein eq})
rewrite (including the contribution of the ordinary matter,
$T^{(m)}_{\alpha\beta}$) $$ \Phi\,R_{\alpha\beta} \,\,-
\,\,\frac{1}{2}\,g_{\alpha\beta} \,{\nabla}_{\nu}\,{\nabla}^{\nu}\,\Phi \,\,-
\,\,{\nabla}_{\alpha}\,{\nabla}_{\beta}\,\Phi = \frac{8\pi G}{c^{4}} \,[
\,T^{(EM)}_{\alpha\beta} \,\,+ \,\,(\, T^{(\psi)}_{\alpha\beta} \,\,-
\,\,\frac{1}{2} \,T^{(\psi)}\,g_{\alpha\beta} \,) $$ \begin{equation}
\label{Einstein eq bis} \,\,+ \,\,(\, T^{(m)}_{\alpha\beta} \,\,-
\,\,\frac{1}{2} \,T^{(m)}\,g_{\alpha\beta} \,) \,].  \end{equation} Now in the
weak fields and slow motion approximation, only the $00$ components are
relevant.  Hence, the above equations reduce to \begin{equation} \label{Einstein
eq approx} \Delta ( \Phi\,g_{00} ) =  \vec{\nabla} \Phi \,. \,\vec{\nabla}
\ln{\sqrt{-\,g}} \,\,+ \,\,\frac{8\pi G}{c^{4}} \,[ \,T^{(EM)}_{00} \,\,+ \,\,(\, T^{(\psi)}_{00}
\,\,- \,\,\frac{1}{2} \,T^{(\psi)}\,g_{00} \,) \,\,+ \,\,(\, T^{(m)}_{00} \,\,-
\,\,\frac{1}{2} \,T^{(m)}\,g_{00} \,) \,].  \end{equation}  The first
term of the right hand side of the above equation is second order. Hence, neglecting the
energy densities of the EM field and the $\psi$-field with respect to the
(ordinary) matter density, one gets \begin{equation} \label{Einstein eq weak
fields and slow motion approx} \Delta ( \Phi\,g_{00} ) = \frac{8\pi
G}{c^{4}} \,(\, T^{(m)}_{00} \,\,- \,\,\frac{1}{2} \,T^{(m)}\,g_{00} \,).
\end{equation} Clearly, this yields the Newtonian potential divided by $\Phi$,
that is an effective potential where $G_{N}$ is replaced by $G_{eff} = G/\Phi$.

\section{Appendix B} As a consequence, the equation of motion of a satellite
should be sensitive to the effect of the effective G if present.  Indeed, in the
static gravitational field of a rotating body of mass $M$ with angular velocity
$\omega$ and for $r \simeq 2a$ or more, one may merely replace in the first
approximation $G_{N}$ by $G_{eff}$ in the Newtonian potential.  Thus,
$d^{2}\vec{r}/dt^{2} = - \,\vec{\nabla} V = - \,( \,g_{r}\,{\vec{u}}_{r} \,+
\,g_{\theta}\,{\vec{u}}_{\theta} \,)$, where ${\vec{u}}_{r} = \vec{r}/r$,
${\vec{u}}_{\theta} = d{\vec{u}}_{r}/d\theta$ and \begin{equation}
\label{effective potential} V = -\,\frac{G_{eff}\,M}{r^{2}} \,[ \,1 \,- \,(
\,\frac{a}{r} \, )^{2} \,J_{2}\,( \,\frac{3}{2} \,cos^{2}{\theta} \,-
\,\frac{1}{2} \,) \,] \,- \,{\omega}^{2} \,r \,( \,1 \,- \,cos^{2}{\theta} \,).
\end{equation} Hence, inserting relations (\ref{stabilized KK scalar sol.  stat.
1st approx} - \ref{def.  mixed variable x}) in equation (\ref{effective g})
above and expanding, one gets $$ g_{r} = \frac{G'(r)\,M}{r^{2}} \,\{ \,1 \,- \,3
\,( \,\frac{a}{r} \, )^{2} \,J'_{2}\,( \,\frac{3}{2} \,cos^{2}{\theta} \,-
\,\frac{1}{2} \,) \,+ \,k(r) \,\epsilon(\theta, \varphi) \,[ \,1 \,- \,3 \,(
\,\frac{a}{r} \, )^{2} \,J_{2}\,( \,\frac{3}{2} \,cos^{2}{\theta} \,-
\,\frac{1}{2} \,) \,] $$ \begin{equation} \label{effective g} \,- \,\frac{9}{2}
\,k(r) \,( \,\frac{a}{r} \, )^{2} \,J_{2} \,cos^{4}{\theta} \,\} \,-
\,{\omega}^{2} \,r \,( \,1 \,- \,cos^{2}{\theta} \,) \end{equation} and
analogously for $g_{\theta}$, where \begin{equation} \label{redefined J2} J'_{2}
= \frac{J_{2}\,( \,1 \,- \,\frac{13}{3} \,k(r) \,) \,- \,\frac{2}{9}\,k(r) \,(
\,\frac{r}{a} \,)^{2}}{1 \,+ \,\frac{k(r)}{3} \,[ \,13 \,+ \,\frac{3}{2} \,(
\,\frac{a}{r} \,)^{2} \,J_{2} \,]} \end{equation} appears as the effective
quadrupole moment coefficient of the central body at radius $r$.  The quadratic
cosine term provides an additional term to the effective zonal harmonic
coefficient of order $4$.  Clearly, by interpreting the data from a single
satellite in circular orbit, $J'_{2}$ will appears as a constant parameter.

\end{document}